\documentclass[11pt,journal,twoside]{IEEEtran}
\usepackage{graphicx}
\usepackage{epstopdf}
\usepackage{placeins}
\usepackage{array}
\DeclareGraphicsExtensions{.eps,.pdf} 
\graphicspath{ {figures/} }
\usepackage{cite}
\usepackage{balance}
\usepackage{subfigure}
\usepackage[ruled,vlined]{algorithm2e}
\usepackage{amssymb,amsmath,mathtools}
\usepackage{amssymb,amsmath}
\usepackage{algorithmic}
\usepackage{color}
\usepackage{multirow}
\usepackage{stackrel,amssymb,amsmath}
\usepackage{empheq}
\usepackage{cases}
\DeclareMathAlphabet{\mathpzc}{OT1}{pzc}{m}{it}

\makeatletter
\newcommand\restartchapters{\par
  \setcounter{chapter}{0}%
  \setcounter{section}{0}%
  \gdef\@chapapp{\chaptername}%
  \gdef\thechapter{\@arabic\c@chapter}}
\makeatother

\def\bq{{\bf q}}

\def\bq{{\bf q}}

\newtheorem{definition}{\bf Definition}

\makeatletter
\g@addto@macro\normalsize{%
 \setlength\abovedisplayskip{4pt}
 \setlength\belowdisplayskip{4pt}
 \setlength\abovedisplayshortskip{4pt}
 \setlength\belowdisplayshortskip{4pt}
}
\makeatother

\makeatletter
\def\endthebibliography{%
	\def\@noitemerr{\@latex@warning{Empty `thebibliography' environment}}%
	\endlist
}
\makeatother

\begin{document}
\bstctlcite{IEEEexample:BSTcontrol}
\title{Deep Reinforcement Learning for Network Energy Saving in 6G and Beyond Networks }
\author{ \normalsize
\IEEEauthorblockN{$\text{Dinh-Hieu Tran}$, $\text{Nguyen Van Huynh}, \textit{Member, IEEE}$, $\text{Soumeya Kaada}$, $\text{Van Nhan Vo}$, $\text{Eva Lagunas}, \textit{Senior, IEEE}$ $\text{and Symeon Chatzinotas}, \textit{Fellow Member, IEEE},$  }
\thanks{Dinh-Hieu Tran, Eva Lagunas, and Symeon Chatzinotas, are with the $\&$ Interdisciplinary Centre for Security, Reliability and Trust (SnT), the University of Luxembourg, Luxembourg. (e-mail: \{hieu.tran-dinh, eva.lagunas, symeon.chatzinotas \} @uni.lu).}
\thanks{N. V. Huynh is with the Department of Electrical Engineering and Electronics,
	University of Liverpool, Liverpool, L69 3GJ, United Kingdom (e-mail:
	huynh.nguyen@liverpool.ac.uk).}
\thanks{S. KAADA is with the Department of computer science, University of Rennes 1- Beaulieu Campus, 35000 Rennes, France (e-mail:
	soumeya.kaada@etudiant.univ-rennes1.fr).}
\thanks{V. N. Vo is with the Faculty of Information Technology, Duy Tan University, Da Nang 550000, Vietnam, and the Institute of Research and Development, Duy Tan University, Da Nang 550000, Vietnam (e-mail: vonhanvan@dtu.edu.vn.}
}
\maketitle
\thispagestyle{empty}
\pagestyle{empty}

\begin{abstract}
Network energy saving has received great attention from operators and vendors to reduce energy consumption and CO2 emissions to the environment as well as significantly reduce costs for mobile network operators. However, the design of energy-saving networks also needs to ensure the mobile users' (MUs) QoS requirements such as throughput requirements (TR). This work considers a mobile cellular network including many ground base stations (GBSs), and some GBSs are intentionally turned off due to network energy saving (NES) or crash, so the MUs located in these outage GBSs are not served in time. Based on this observation, we propose the problem of maximizing the total achievable throughput in the network by optimizing the GBSs' antenna tilt and adaptive transmission power with a given number of served MUs satisfied. Notice that, the MU is considered successfully served if its Reference Signal Received Power (RSRP) and throughput requirement are satisfied. The formulated optimization problem becomes difficult to solve with multiple binary variables and non-convex constraints along with random throughput requirements and random placement of MUs. We propose a Deep Q-learning-based algorithm to help the network learn the uncertainty and dynamics of the transmission environment. Extensive simulation results show that our proposed algorithm achieves much better performance than the benchmark schemes.
\end{abstract}

\begin{IEEEkeywords}
6G, Deep Q-Network, Network Energy Saving, Power Saving, Self-Organizing Networks (SONs).
\end{IEEEkeywords}

\label{sec:Intro}
Energy consumption (EC) has become a significant part of an operator’s operating expenses or expenditures (OPEX). The EC on mobile networks accounts for ~23$\%$  of the total operator cost \cite{MariaNES}. Most of the power consumption comes from the radio access network (RAN) and especially from the Active Antenna Unit (AAU), with data centers and fiber optic transmission accounting for a smaller share. The power consumption (PC) of RAN can be split into two parts: the dynamic part which is only consumed when data transmission/reception (Tx/Rx) is ongoing, and the static part which is consumed all the time to maintain the necessary operation of the radio access devices, even when the data Tx/Rx is not taking place.        

Therefore, network energy saving (NES) has received significant attention from vendors and operators\cite{EricssonNES, HuaweiNES}. That becomes even more stringent in 6G and beyond networks (6GBNs) with applications requiring higher speeds and lower latency, higher reliability than 5G, (e.g. extended reality/virtual reality, holographic, 3D video streaming, metaverse). Moreover, 6GBNs are becoming denser, and more antennas, bandwidth, and spectrum to adapt to mobile users' (MUs) QoS requirements. As a result, the environmental impact of 6G needs to be managed and new solutions need to be developed to improve network energy efficiency.     






Recently, self-organizing networks (SONs) using Artificial intelligence/machine learning (AI/ML) have emerged as an effective solution that helps to reduce the GBSs's EC \cite{EricssonMLNES, NokiaMLNES}. With the increasing number of GBSs, mobile users (MUs), and more energy-consuming applications, more EC is inevitable. Manual management has become obsolete and inefficient in managing a large number of GBSs, and MUs for applications with heterogeneous latency, bandwidth, and rate requirements. Therefore, using an automatic operating system that uses machine learning is becoming urgent. Ericsson has shown that AI/ML can dynamically learn and then predict the MUs' traffic demand in specific areas at specific times \cite{EricssonMLNES}. Besides Ericsson, Nokia also prioritizes the use of AI/ML to optimize energy savings at GBSs and reduce Co2 emissions \cite{NokiaMLNES}. Nokia shows that AI/ML can provide effective energy-saving solutions without significantly affecting the user experience. Therefore, AIML has become an indispensable tool for designing an automated system to operate a large number of GBSs and MUs efficiently and energy-efficient manner.

Recently, there have been some studies done to apply AI/ML to NES \cite{bassoy2023seedrl,zhou2022hierarchical,wu2023energy, lopez2024data}. Selcuk et al. \cite{bassoy2023seedrl} proposed a new NES algorithm called Smart Energy
Efficiency using Deep Reinforcement Learning (SEEDRL) which can save up to around 9$\%$ power compared to the literature. More specifically, their algorithm aims to turn off some specific GBSs and antenna elements at each GBS to save energy without affecting the MUs's QoS. Zhou et al. \cite{zhou2022hierarchical} applied intelligent reflecting surface (IRS) as a promising technology to save energy in 6GBNs. In their study, they designed a hierarchical reinforcement learning (HRL) algorithm to control the GBSs at two levels, i.e., turn on/off the GBSs and change the transmit power of each GBSs for NES purposes. Moreover, IRS was used in the system model to increase the average throughput of the network. In \cite{wu2023energy}, Wu et al. applied transfer DRL (TDRL) to improve the NES in 6GBNs. Their purpose was to develop a novel TDRL algorithm in NES for better data efficiency compared to the baseline DRL approach. In \cite{lopez2024data}, David et al. proposed a novel simulated reality of communication networks (SRCON) framework that exploited real-time network data and deployed multiple ML network models for NES. Compared to other simulators used by other operators, SRCON showed the advantages in 63$\%$ higher accuracy of the mean absolute error (MAE) and mean absolute percentage error (MAPE). 

Although they give many interesting results, none of the above studies \cite{bassoy2023seedrl,zhou2022hierarchical,wu2023energy, lopez2024data} mention solutions when AI/ML makes the wrong decision to turn off one or multiple GBSs \cite{rich2019lessons}. AI/ML can help us analyze the data from the past and help us to make good decision making in the future. However, with the constant change of the transmission environment in telecommunication, such as traffic demands, channel conditions, and mobility, AI/ML cannot guarantee 100$\%$ that all of its decisions are completely correct \cite{rich2019lessons}. In this case, how do we cover the existing UEs and guarantee an acceptable performance before the network realizes it and turns on GBSs again. In NES, turning off GBSs to save more energy during off-peak hours is necessary but only if the MUs' QoS requirement is met. Motivated by these observations, we developed this study to ensure network robustness when wrong decision-making happens in NES for 6GBNs. Our contributions are summarized as follows:
\begin{itemize}
	\item To the best of our knowledge, this is the first work that investigates the time domain (i.e., turn-off GBSs), power domain (i.e., GBSs' power adaptation), space domain (i.e., GBSs' antenna tilt adaptation) by applying DRL  ensuring the robustness of 6GBNs when wrong decision-making to turn off GBSs happens by scheduling surrounding GBSs to support as much as possible the MUs distributed in the inactive areas. Especially, each MU requests different QoS requirements (i.e., throughput) and is located randomly around the GBS. 
	\item We consider a multi-cell RAN network that considers the GBSs inactive/turned-off to save energy, random throughput demands from MUs, and random distribution of MUs's position. We formulate an optimization problem to maximize the achievable sum rate under the maximum allocated bandwidth, minimum and maximum GBS's transmit power, reference signal received power (RSRP) constraint, minimum MU's throughput requirement, and with a given number of successfully served MUs. The formulated problem belongs to mixed-integer non-linear programming (MINLP) which is NP-hard. 
	\item To obtain the optimal policy under random throughput demand of MUs and random MUs's location, we deploy the DQN-based algorithm that can obtain the optimal solution through the reinforcement learning process. The proposed DQN-based algorithm is presented in the simulation results, which show the superiority compared to the benchmark ones.
\end{itemize}

\begin{figure}[t]
	\centering
	\includegraphics[height=6cm,width = 8.5cm]{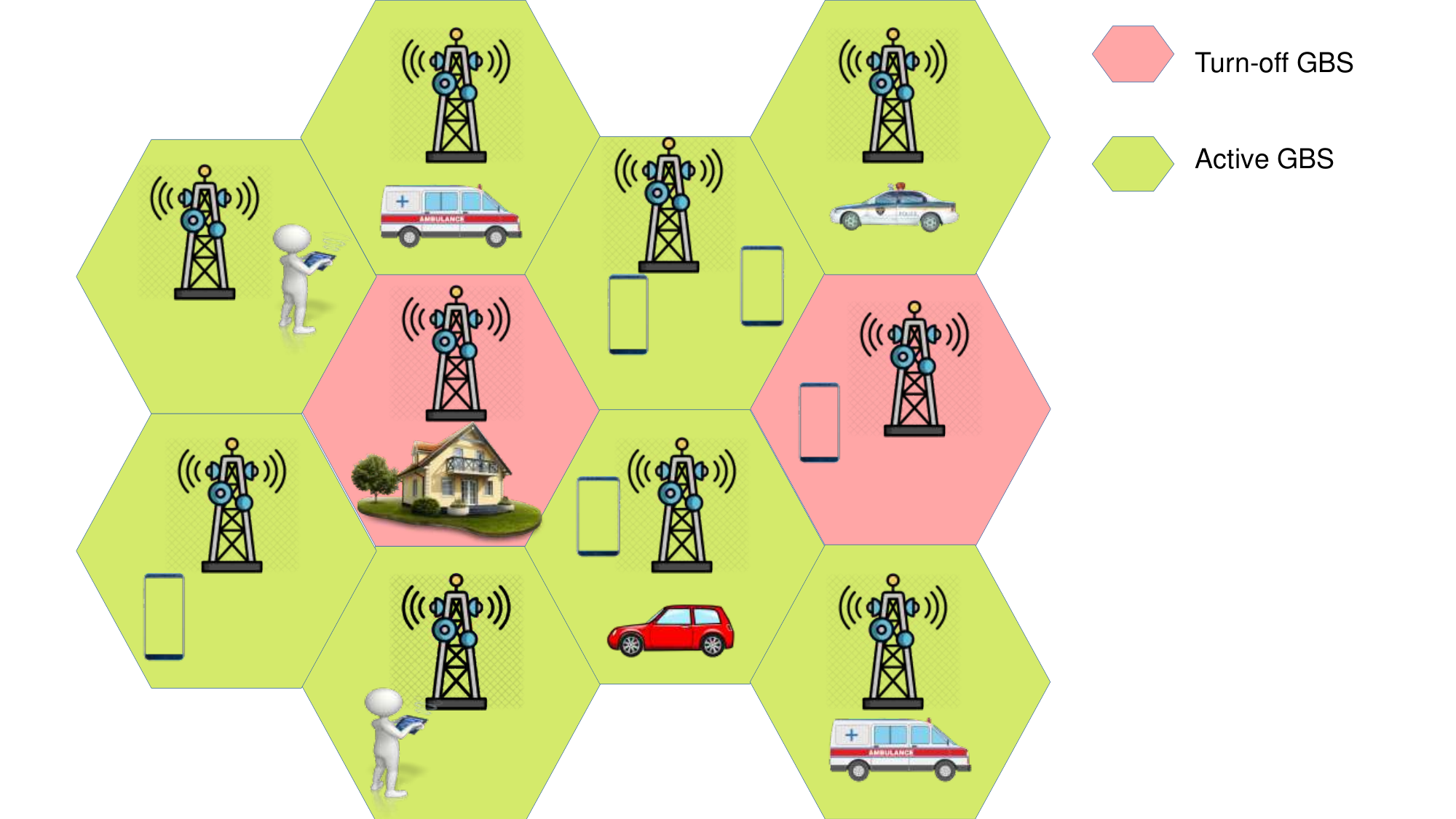}
	\caption{System model.}
	\label{Fig1}
\end{figure}

\section{System Model}
\label{sec:Sys}

In Fig. \ref{Fig1}, we consider a downlink (DL) multi-cell system, in which the GBS is located at the center of each cell. The set of $K$ GBSs and the set of all $U$ MUs distributed in these cells are denoted by ${\mathcal{K}} = \{1, \dots,k, \dots, K\}$ and ${\mathcal{U}} = \{1, \dots,u, \dots, U\}$, respectively. Moreover, each GBS has three different antennas to cover three sectors, the set of antenna $i$ of each GBS is denoted by $i \in {\mathcal{I}} = \{1,2,3 \}$. 

\subsection{Channel Model}
Let us define the distance between GBS $k$ to MU $u$ in 3D is expressed as:
\begin{align}
	\label{eq:2}
	d_{uk}^{\rm 3D} &= \sqrt{(h_k-h_u)^2 + || \bq_k - \bq_u ||^2},
\end{align}
where $h_k$ and $h_u$ are the altitude of the GBS $k$ and MU $u$, respectively. Moreover, we use the Cartesian coordinates to define the location of GBS $k$ and MU $u$, respectively, i.e.,  $\boldsymbol{q}_j \in {\mathbb{R}}^{2 \times 1}$ or $\boldsymbol{q}=[x_j,y_j]$ with $j \in \{k,u\}$.

In this work, we consider a practical system model including both light-of-sight (LoS)
and non-light-of-sight (NLoS) paths that adapt to different environments, i.e., urban or rural areas. Specifically, we consider both large-scale and small-scale fading \cite{HieuTVT22, HieuOJVT22}. Then, when the GBS $k$ transmits a signal with transmit power $P_k^i$ through antenna $i$ $\rightarrow$ the MU $u$ with the effects of large-scale path loss, small-scale Rician fading, and antenna gains. Then, the received power at the MU $u$ is given as \cite{wang2018connectivity,Dung2016does}:

\begin{align}
	\label{eq:9}
	{\mathcal P}_{uk}^i = {\mathcal P}_k^i  |h_{uk}^i|^2  (d_{uk}^{\rm 3D})^{-\alpha} {\mathfrak G_k^i(\theta, \psi)}  G_u  , 
\end{align}
where ${\mathcal P}_k^i$ is the transmit power the the GBS via antenna $i$, $h_{uk}^i$ is the small-fading Rician fading, $d_{uk}^{\rm 3D}$ is the 3D distance from GBS $k$ $\rightarrow$ the MU $u$,$G_u$ is the received antenna gain at the MU $u$, and ${\mathfrak G_k^i(\phi, \theta)}$ denotes the transmit antenna gain of the GBS $k$ via antenna $i$ which is defined in the next subsection antenna model \ref{subsec_Antenna}.

Based on \eqref{eq:9}, the throughput (in Bits/s/Hz) at the MU $u$ when receives the transmitted signal from GBS $k$ is given as:
\begin{align}
	\label{eq:throughput}
	{\mathcal R}_{uk}^i &=   \log_2 \left(1+\Gamma_{uk}\right),
\end{align}
with:
\begin{align}
	\label{eq:throughput}
	\Gamma_{uk} \triangleq \frac{ {\mathcal P}_{uk}^i } { ((h_k-h_u)^2 + || \bq_k - \bq_u ||^2)^{\alpha/2} \big(\phi^{\rm RIC} \sum\limits_{k^\ast \in {\cal K} \setminus k} {\mathcal P}_{uk^\ast}^i +  \sigma^2\big) },
\end{align}
where $\Phi^{\rm RIC} \triangleq \phi^{\rm RIC} \sum\limits_{k^\ast \in {\cal K} \setminus k} {\mathcal P}_{uk^\ast}^i$ denotes the residual interference term after canceling all the interference from other GBS $k^\ast \in {\mathcal{K}} \setminus k$ \cite{tran2021uav}. $\sigma^2$ is the variance of the additive white Gaussian noise (AWGN) $n_0$ with $n_0 \sim {\cal{CN}}(0,\sigma^2)$.

Since $h_{uk}^i$ is a random variable (RV), the throughput ${\mathcal R}_{uk}^i$ is also a RV. Therefore, we aim to obtain the approximated throughput (in Bits/s/Hz) as follows \cite{tran2021uav}:
\begin{align}
	\label{eq:Lemma1_1}
	\bar{\mathcal R}_{uk}^i = \log_2 \Bigg(1+\frac{e^{-E} \widehat{\mathcal P}_{uk}^i   } { {( (h_k-h_u)^2 + {{\left\| {\bq_k - \bq_u} \right\|}^2})^{\alpha/2}} \nu_{1k}^i }\Bigg),
\end{align}
where $\widehat{\mathcal P}_{uk}^i = {\mathcal P}_k^i * (d_{uk}^{\rm 3D})^{-\alpha} * {\mathfrak G_k^i(\theta, \psi)} * G_u  $, $\nu_{uk}^i \triangleq \big(\phi^{\rm RIC} \sum\limits_{k^\ast \in {\cal K} \setminus k} {\mathcal P}_{uk^\ast}^i+  \sigma^2\big)$, $E$ is the Euler-Mascheroni constant, i.e., $E=0.5772156649$ as in \cite[Eq. 8.367.1]{gradshteyn2014}.

\subsection{Antenna Model}
\label{subsec_Antenna}
In this work, we utilize a practical 3D antenna pattern defined by 3GPP \cite{R1083070, TR36814}. First, we calculate the elevation angle $\psi$ and azimuth angle $\theta$ between GBS $k$ $\rightarrow$ MU $u$ in elevation/vertical and azimuth/horizontal planes, respectively, as follows \cite{R1083070, TR36814,viering2009mathematical}:
\begin{align}
	\label{eq:angle}
	\theta_{uk} & = {\rm arctan} \left( \frac{h_k-h_u}{d_{uk}^{\rm 2D}} \right), \\
	\psi_{uk} &= {\rm arctan} \left( \theta_u - \theta_k \right),
\end{align}
where $\theta_u$ and $\theta_k$ are the azimuth angle from the x-axis $\rightarrow$ the main lobe's center of GBS $k$ and from the x-axis $\rightarrow$ MU $u$, respectively. Moreover, the $\theta_k$ values are $0^0$, $120^0$, and $-120^0$ corresponding to three antennas in three sectors \cite{viering2009mathematical}.

As defined in \cite{R1083070, TR36814}, the azimuth and elevation gain models are respectively given as:
\begin{align}
	\label{azimuthgain}
	{\mathfrak G_k^i(\psi_{uk}) }= -\min \left\{12* \left(\frac{\psi_{uk}}{\psi_{\rm 3dB}} \right)^2, {\mathtt F} \right\} + {\mathfrak G_{\max}}, \\ \label{elevationgain}
	{\mathfrak G_k^i(\theta_{uk}) }= -\min \left\{12* \left(\frac{\theta_{uk}- \theta_{uk}^{\rm tilt} }{\theta_{\rm 3dB}} \right)^2, {\mathfrak F} \right\} + {\mathpzc S },
\end{align}
where ${\mathfrak G_{\max}}$ = 14 dBI is the maximum gain; $\psi_{\rm 3dB}$ is the azimuth half-power beamwidth and equals $70^0$;  ${\mathtt F}$ is the front back ratio and equals 20 dB; $\theta_{uk}^{\rm tilt}$ denotes the antenna down-tilt angle including both mechanical (i.e.,$\theta_{uk}^{\rm mtilt}$) and electrical down-tilt (i.e.,$\theta_{uk}^{\rm etilt}$), when this value changes it will influence to both $\theta_{uk}$ and $\psi_{uk}$ values; $\theta_{\rm 3dB}$ and $\theta_{\rm 3dB}$ denote the azimuth and elevation half-power beamwidth and can be set to $65^0$ and $65^0$, respectively; 

Since Eqs. \eqref{azimuthgain} and \eqref{elevationgain} represent the antenna gain along the horizontal and vertical, respectively. Therefore, to describe the antenna gain 3D pattern, we need to combine these two antenna gain patterns \cite{R1083070} as follows:
\begin{align}
	\label{3Dgain}
	{\mathfrak G_k^i(\theta_{uk}, \psi_{uk})} = {\mathfrak G_k^i(\theta_{uk})} + {\mathfrak G_k^i(\psi_{uk})}.
\end{align}

\section{Network Energy Saving Problem}
\label{sec:Problem}
In this section, we describe how we formulate the optimization problem. In our system, when one or more GBSs are off, the surrounding GBSs need to cover the MUs in these off GBSs. However, due to the limited resources, not all the MUs are guaranteed their QoS, e.g., the RSRP or throughput requirement. 

Due to energy saving, a GBS $k$ can be turned off, then it yields:
\begin{subnumcases} 
	{\chi_{k}  =}
	1, \hfill \text{GBS} \; k \; \text{is active}, \label{eq:RSRPa}\\
	0, \hfill \text{GBS} \; k \; \text{is off}.  \label{eq:RSRPb}
\end{subnumcases}

Inspired by a practical wireless communication system, the MUs can be served from GBS $k$ if it RSRP is satisfied, i.e.,:
\begin{subnumcases} 
	{\vartheta_{uk}  =}
	1, \hfill P_{uk} \ge P_{uk}^{\rm th} \label{eq:RSRPa}\\
	0, \hfill P_{uk} \le P_{uk}^{\rm th}  \label{eq:RSRPb}
\end{subnumcases}

Then, if the MU $u$ satisfies the RSRP condition, we add one more condition to satisfy MU $u$'s throughput requirements as follows:
\begin{subnumcases} 
	{\gamma_{uk}  =}
	1, \hfill {\mathcal R}_{uk}^i \ge {\mathcal R}_{uk}^{\rm th} \label{eq:RSRPa}\\
	0, \hfill {\mathcal R}_{uk}^i \le {\mathcal R}_{uk}^{\rm th}  \label{eq:RSRPb}
\end{subnumcases}

\begin{definition}
	The MU $u$ called successfully served iff its RSRP and throughput constraints are satisfied. 
\end{definition} 

Then, we introduce a new binary variable $\pi$ such that:
\begin{subnumcases} 
	{\pi_{uk}  =}
	1, \hfill \vartheta_{uk}   = 1 \; $\&$ \; \gamma_{uk}  = 1,\label{eq:RSRPa}\\
	0, \hfill {\rm Otherwise}. \label{eq:RSRPb}
\end{subnumcases}

Let us define ${\boldsymbol \vartheta} \triangleq \{\vartheta_{uk}, \forall u,k\}$, ${\boldsymbol \gamma} \triangleq \{\gamma_{uk}, k \in {\cal K}, u \in {\cal U} \},$  ${ \boldsymbol \pi} \triangleq \{{\pi_{uk}}, k \in {\cal K} , u \in {\cal U}\}$, ${ \boldsymbol \theta} \triangleq \{\theta_{uk}^{\rm tilt}, k \in {\cal K} , u \in {\cal U}\}$, ${ \boldsymbol {\mathcal P}} \triangleq {\mathcal P}_k^i, \forall k, i$. Based on the above discussion, the problem of maximizing the achievable throughput at all MUs with a given number of served MUs satisfied is subject to RSRP and QoS constraints given as follows:

\begin{align}
	\label{eq:P1}
	{\cal P}:\ &\max_{{\boldsymbol \vartheta}, {\boldsymbol \gamma}, { \boldsymbol \pi}, {\boldsymbol \theta}, { \boldsymbol {\mathcal P}} }~~ \sum\limits_{k \in {\cal K}, u \in {\cal U} }   \notag \\  &  \log_2 \Bigg(1+\frac{e^{-E} \widehat{\mathcal P}_{uk}^i   } { {( (h_k-h_u)^2 + {{\left\| {\bq_k - \bq_u} \right\|}^2})^{\alpha/2}} \nu_{1k}^i }\Bigg) \\
	\textbf{s.t.:} \notag \\
	& \{\chi_{k}, \vartheta_{uk}, \gamma_{uk}, \pi_{uk} \} \in \{0,1\}, \forall u,k, \IEEEyessubnumber\label{eq:P1:b}\\
	& \left\|  {\boldsymbol \pi}  \right\|_1  \ge \pi_{\rm thresh}, \forall u,k, \IEEEyessubnumber\label{eq:P1:c}
	\\
	\vspace{-0.01cm}
	&  \sum\limits_{ u \in {\cal U} } 
	{\pi}_{uk}    \le \pi_{k}^{\max}, \forall u,k, \IEEEyessubnumber\label{eq:P1:d}
	\\
	& \log_2 \Bigg(1+\frac{e^{-E} \widehat{\mathcal P}_{uk}^i   } { {( (h_k-h_u)^2 + {{\left\| {\bq_k - \bq_u} \right\|}^2})^{\alpha/2}} \nu_{1k}^i }\Bigg) \ge {\mathcal R}_{uk}^{\rm th} , \notag \\ & \forall u,k, \IEEEyessubnumber\label{eq:P1:e}\\
	& {\mathcal R}_{uk}^{\min}  \le {\mathcal R}_{uk}^{\rm th} \le {\mathcal R}_{uk}^{\max}, \forall u,k, \IEEEyessubnumber\label{eq:P1:f}\\
	& {\mathcal P}_{k}^{\min}  \le {\mathcal P}_{k}^i \le {\mathcal P}_k^{\max}, \forall u,k, \IEEEyessubnumber\label{eq:P1:g}\\
	& d^{\min}  \le {d}_{uk}^{\rm 3D} \le {d}^{\max}, \forall u,k, \IEEEyessubnumber\label{eq:P1:h}\\
	& \theta_{uk}^{\rm tilt} \in \left[0^o,14^o\right], \; \forall u,k, \IEEEyessubnumber\label{eq:P1:i}
\end{align}
where \eqref{eq:P1:c} represents the number of served MUs that should be bigger than or equal to a threshold, i.e., $\pi_{\rm thresh}$; Eq. \eqref{eq:P1:d} shows the maximum number of served MUs in one GBS is limited by a maximum value $\pi_k^{\max}$ representing for the limited resources at that GBS; Eq. \eqref{eq:P1:e} denotes the rate requirement for each MU need to be higher than or equal to a threshold value, i.e., ${\mathcal R}_{uk}^{\rm th}$; Eq. \eqref{eq:P1:f} shows the heterogeneous traffic demand of each MU; Eq. \eqref{eq:P1:g} represents the minimum and maximum transmit power of a GBS $k$; Eq. \eqref{eq:P1:h} denotes the random locations of MUs with different distance to the GBS; Eq. \eqref{eq:P1:i} represents the range values of antenna tilt in one GBS.

The problem $\cal P$ is a mix-integer nonlinear problem (MINLP), which is NP-hard. In the following section, we propose an efficient solution for applying DQN. 

\section{Deep Q-Network-based Solution}
\label{sec:DQN}

To overcome the uncertainty of the heterogeneous on-off GBS, MUs' traffic demand, and positions, we formulate the NES as a Markov Decision Process (MDP) \cite{hoang2023deep}. More specifically, an MDP is described by a tuple $\Big(\mathcal{A}, \mathcal{S}, {\mathcal{R}}\Big)$m where $\mathcal{A}, \mathcal{S}, $ and ${\mathcal{R}}$ are state space, action space, and immediate reward, respectively. In the following, we describe the state space, action space, and reward function. A policy $\mathcal{\pi}$ showing the probability of selecting an action $a$ in state $s$, and can be given as $\mathcal{\pi}(a|s)$.
\subsection{State}
\label{statespace}
It represents the configuration of the antenna tilt angle (ATA) and transmit power level of each antenna belonging to a GBS. The state can be modeled by an M x 2 matrix where M denotes the number of antennas in each GBS and 2 represents the power level and antenna tilt of each antenna.
\subsection{Action Space}
\label{Actionspace}
In each sector/cell, we have three different actions for adapting the ATA, i.e., $\{ -1^0, 0^0, +1^0 \}$, and three different actions for changing the GBS transmit power, i.e., $\{ -5 \;{\rm dB}, 0 \;{\rm dB}, +5 \; {\rm dB} \}$. Therefore, we have a total of nine actions per sector/cell. For a GBS that includes three sectors, we have $9^3$ actions. Then, we have an action space of $9^{\mathcal C}$ if we consider an area of $\mathcal C$ cells.

\subsection{Immediate Reward}
\label{Actionspace}
${\mathcal{R}}: {\mathcal{S}} \times {\mathcal{A}}\times{\mathcal{S}}$ represent the reward function, which shows the objective to optimize in \eqref{eq:P1}. In this case, we consider the sum rate maximization of all MUs in the networks. Then, the reward function at state $s$ and with action $a$ is defined as follows:
\begin{subnumcases}
	{{\mathcal R}_{s,a,s'}  =}
	\sum\limits_{k \in {\cal K}, u \in {\cal U} }  \bar{\mathcal R}_{uk}^i , \hfill \vartheta_{uk}   = 1 \; $\&$ \; \gamma_{uk}  = 1,\label{eq:rewarda}\\
	0, \hfill {\text{Otherwise} }. \label{eq:rewardb}
\end{subnumcases}

We note that Eq. \eqref{eq:rewarda} plays as a penalty in the learning process that forces our optimization problem to satisfy the RSRP and throughput constraints. 

\subsection{DQN-based Solution}
\subsubsection{Q-learning}
Before talking about DQN, we would like to introduce you to Q-learning, a basic RL algorithm. 

First, we introduce the state value which is the total expected reward value (ERV) that we can have at a specific state:
\begin{align}
	\label{valuefunc}
	V^{\mathcal{\pi}}(s)= {\mathbb{E}} \left[  {\mathcal R}_{t}|s_t=s\right], 
\end{align}

Then, we discuss the value function, which is the ERV that we can have after taking a specific action $a$ at a specific state $s$:
\begin{align}
	\label{valuefunc}
	Q^{\mathcal{\pi}}(s,a)= {\mathbb{E}} \left[  {\mathcal R}_{t}|s_t=s,a_t=a\right], 
\end{align}

The action value function in Q-learning aims to provide a long-term reward, which is given as: \cite{fan2020theoretical}:
\begin{align}
	\label{valuefunc}
	Q^{\mathcal{\pi}}(s_t,a_t)= {\mathbb{E}} \left[ \sum\limits_{t =0 }^\infty \zeta \times {\mathcal R}_{t}|s_0=s,a_0=a\right], 
\end{align}
where $\zeta \in (0,1]$ denotes the discount factor in which $\zeta$ closes to zero which means that the agent will prioritize immediate rewards and $\zeta$ closes to one which means that the agent will prioritize the long-term reward. The objective of Q-learning is to maximize the expected future reward by finding the optimal policy, i.e., $\pi^*$, the agent will learn to make a specific action corresponding to a specific state and iteratively converge to the optimal action-value function. As a result, the optimal $Q^{\mathcal{\pi}^\star}(s_t,a_t)$ function can be represented mathematically by applying the Bellman equation as:
\begin{align}
	\label{valuefunc}
	Q^{\mathcal{\pi}^*}(s_t,a_t)={\mathcal R} (s_t,a_t) +\zeta  {\mathbb{E}} \left[  \sum\limits_{a_{t+1}} Q_{s_{t+1},a_{t+1}}|s,a\right], 
\end{align}
\subsubsection{Deep Q-Network}
In DQN, we use the deep neural network (DNN) $Q(s_t,a_t;\omega)$ to obtain the optimal value of $Q^{\mathcal{\pi}^*}(s_t,a_t)$, where $\omega$ is the parameters of the DDN. The DNN inputs state $s$ and outputs $Q$ value for all actions. It aims to train the NN with its output to the true Q values. The DQN applies the experience relay to store the dataset from prior experience to a buffer. The prior experience is presented as tuples of ${\mathfrak e}_t \triangleq  (s_t,a_t,R_t,s_{t+1})$ into a buffer ${\mathfrak B} \triangleq \{{\mathfrak e}_1, \dots,  {\mathfrak e}_n\}$. Then, DQN samples a random mini-batch from ${\mathfrak B} $ to train the NN. At each iteration, the DQN agent will select an action applying $\epsilon-$greedy policy. Note that the parameter $\omega$ is not updated from the previous iteration $\omega_{i-1}$. We use the parameters from a few iterations ago but do not use the last parameter's iteration \cite{Tensorflow}. After each iteration, the agent will select the best action to maximize the long-term cumulative reward.


\section{Simulation Results}
\label{Results}
In this section, we present simulation results to evaluate the proposed joint deactivated GBSs, RSRP constraint, MU's QoS requirement, as well as the adapted antenna tilt in 6G and beyond network energy saving. We consider a system with $U$ MUs that are randomly distributed in $K$ GBSs following a uniform distribution. Moreover, these MUs demand a different QoS requirement, i.e., rate.  Other parameters are defined as follows: number of MUs $\left\|  {\mathcal U}  \right\|_1 $ = 80 $\rightarrow$ 2000 MUs, the height of the MU is $h_u=1.5$ meters, the height of the GBS is $h_k$ = 10 meters, mechanical down-tilt value is $\theta_{uk}^{\rm mtilt}=0^0$, electrical down-tilt value is $\theta_{uk}^{\rm rtilt}=[0^0, 14^0]$, maximum gain of the antenna ${\mathfrak G_{\max}}=20$ dBi, minimum and maximum transmit power of the GBS are $P_{k}^{\min}=0$ dBm and $P_{k}^{\max}=45$ dB, azimuth and elevation half-power beamwidth and can be set to $65^0$ and $65^0$, respectively, the maximum queue of the buffer in memory relay in DQN is 20000. The first benchmark scheme is maximum-action-based algorithm (a.k.a. Max in the below figures), where the agent always chooses to increase the highest antenna tilt and transmit power of GBS for each action, i.e., $+1^0 \in \{ -1^0, 0^0, +1^0 \}$,   $+5 \; {\rm dB} \in  \{ -5 \;{\rm dB}, 0 \;{\rm dB}, +5 \; {\rm dB} \}$. The second benchmark scheme is a random-based algorithm (a.k.a. Random in the below figures), where the agent always chooses random values of antenna tilt and transmit power of GBS for each action, i.e., $-1^0|0^0|+1^0 \in \{ -1^0, 0^0, +1^0 \}$,   $-5 \;{\rm dB}|0 \;{\rm dB}|+5 \; {\rm dB} \in  \{ -5 \;{\rm dB}, 0 \;{\rm dB}, +5 \; {\rm dB} \}$.

\begin{figure}[t]
	\centering	\includegraphics[height=8cm,width = 9cm]{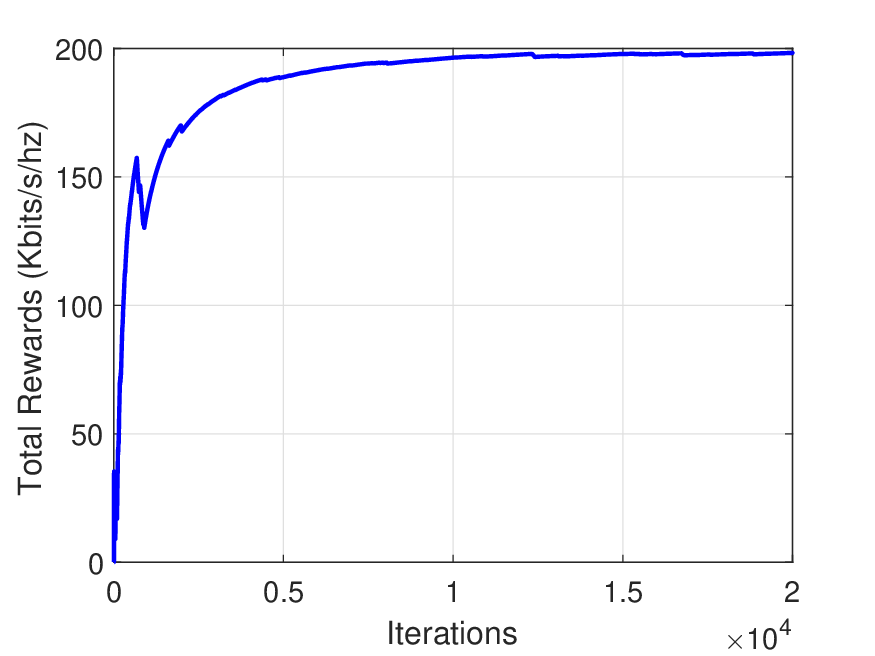}
	\caption{Average rewards vs. number of iterations.}
	\label{Fig3}
\end{figure}

\begin{figure}[t]
	\centering	\includegraphics[height=8cm,width = 9cm]{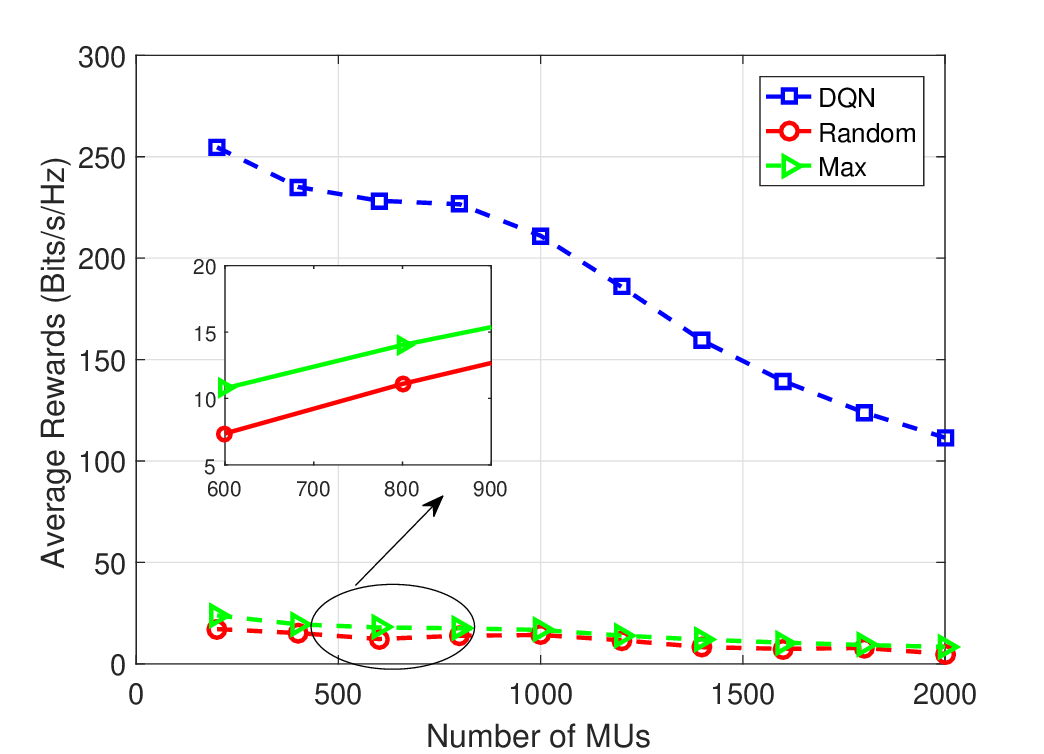}
	\caption{Average reward per MU vs. number of MUs.}
	\label{Fig7}
\end{figure}

In Fig. \ref{Fig3}, we show the convergence average reward of our proposed DQN-based achievable throughput maximization problem with a given number of served MUs when the learning rate is 0.001, the batch size is set to 32 data points, and with 20000 iterations. It can be observed from Fig. \ref{Fig3} that as the increasing of a number of iterations of training, the agent improves the achievable throughput by increasing the reward values. In early iterations, i.e., the number of interactions is less than 2000, we see a large fluctuation in the average reward, it is because the agent is still spending more effort on the exploration phase to obtain the best policy. After that, the average reward increases gradually because the more exploration is applied, the better policy is obtained, which reduces the fluctuation. Another observation is that our proposed model converges to a saturated value after around 11000 iterations when the policy is quite stable. Therefore, it shows the correctness of our proposed DRL-based approach. To illustrate the superiority of our proposed solution, we compared the proposed DQN-based algorithm with benchmark schemes. 



To investigate the influence of different numbers of MUs values on the reward ${\mathcal R}_{s,a,s'}$, Fig. \ref{Fig7} illustrates the average rate of each MU over different MUs numbers, ranging from 200 to 2000. First, it is easy to see that our proposed DQN-based solution achieves a much better sum throughput compared to the Random and Max schemes. This shows the advantages of our proposed DQN-based solution. Second, we observe that the percentage gap between the DQN-based solution to the Random and Max becomes larger with the increasing number of MUs values. For example, when $\left\|  {\mathcal U}  \right\|_1 $ equals 200, the average throughput of the DQN, Random, and Max are 254.67 Bits/s/Hz, 17 Bits/s/Hz, and 23.87 Bits/s/Hz, respectively. It means that the DQN solution performs better at 93.32$\%$ and 90.627$\%$ as compared to the Random and Max schemes, respectively. When $\left\|  {\mathcal U}  \right\|_1 $ equals 2000, the average throughput of the DQN, Random, and Max are 111.52 Bits/s/Hz, 4.925 Bits/s/Hz, and 8.36 Bits/s/Hz, respectively. It means that the DQN solution performs better at 95.58$\%$ and 92.5$\%$ as compared to Random and Max schemes, respectively. It can be explained by the fact that when we select random/maximum values of antenna tilt and transmit power at the GBS, it creates more interference to other MUs. i.e., $\phi^{\rm RIC} \sum\limits_{k^\ast \in {\cal K} \setminus k} {\mathcal P}_{uk^\ast}^i$ in Eq. \eqref{eq:throughput}. Therefore, it will significantly reduce the obtained throughput at each MU and reduce the sum throughput at the objective function in \eqref{eq:P1}. 

\begin{figure}[t]
	\centering
	\includegraphics[height=8cm,width = 9cm]{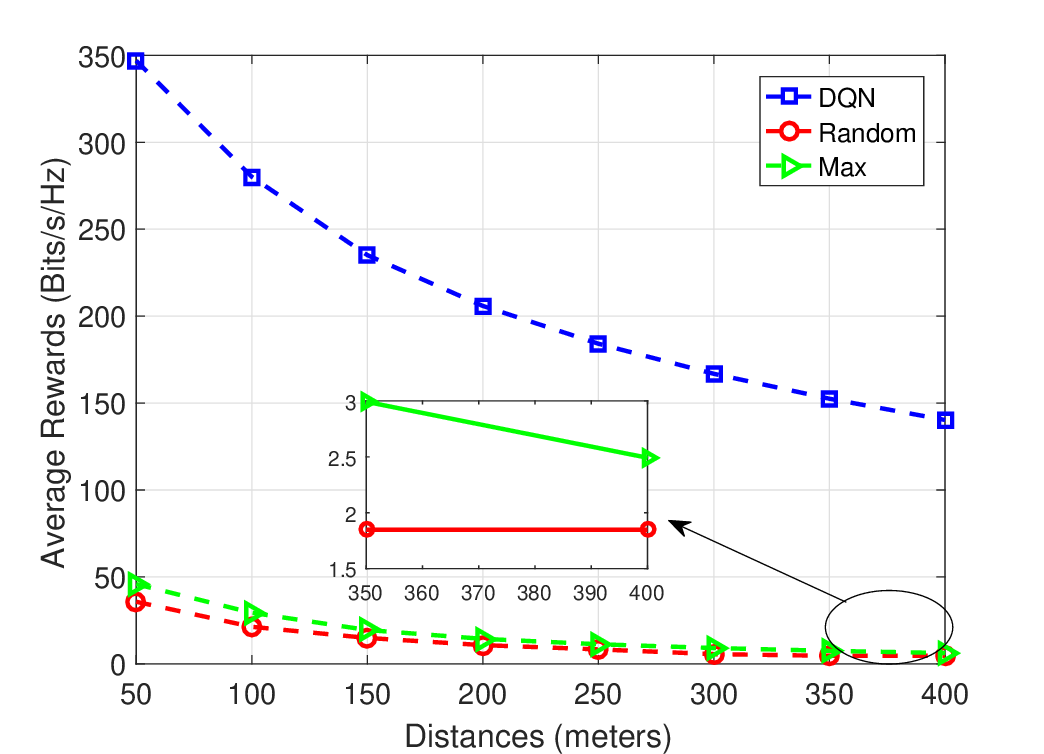}
	\caption{Average reward vs. $d_{uk}^{\rm 3D}$.}
	\label{Fig8}
\end{figure}

In order to show the impact of the average distance from GBS $\xrightarrow{}$ MUs (meters), Fig. \ref{Fig8} shows the sum throughput of all MUs over different average distances from GBS $\xrightarrow{}$ MUs (meters) ranging from 20 meters to 400 meters. More specifically, the distance value ranges from $d^{\rm min}$ $\rightarrow$ $d^{\rm max}$. For simplicity, I only present the $d^{\rm max}$ value in Fig. \ref{Fig8}. For example, when the $d_{uk}^{\rm 3D}$ equals 50, the distance from GBS $\rightarrow$ MUs ranges from 20 meters to 50 meters. Moreover, when the $d_{uk}^{\rm 3D}$ equals 100, the distance from GBS $\rightarrow$ MUs ranges from 50 meters to 100 meters and so on. First, we see that the sum throughput of all MUs is exponentially decreasing with a higher distance from GBS $\xrightarrow{}$ MUs (meters). It is because the distance from GBS $\xrightarrow{}$ MUs (meters) $d_{uk}^{\rm 3D}$ is placed in the denominator in the Eq. \eqref{eq:throughput}. Therefore, when the $d_{uk}^{\rm 3D}$ increases, it will significantly decrease the received throughput ${\mathcal{R}}_{uk}^i$ of each MU $u$, and then influence the sum throughput of all MUs in the networks. Second, we observe that our proposed DQN-based method still outperforms Random and Max schemes. The difference in average reward performance decreases with increasing distance from 30 meters to 200 meters. More specifically, when $d_{uk}^{\rm 3D}$ ranges from 20 $\xrightarrow{}$ 50 meters, the average sum rate of the DQN, Random, and Max are $346.89$ Bits/s/Hz, $35.71$ Bits/s/Hz, and $45.85$ Bits/s/Hz, respectively. It means that the DQN performs better at 89.71$\%$ and 86.78$\%$ compared to the Random and Max schemes, respectively. When $d_{uk}^{\rm 3D}$ ranges from 350 $\xrightarrow{}$ 400 meters, the sum throughput of the DQN, Random, and Max are $140.29$ Bits/s/Hz, $4.62$ Bits/s/Hz, and $6.23$ Bits/s/Hz, respectively. It means that the DQN performs better at 96.71$\%$ and 95.56$\%$, respectively. It is because the higher distance value significantly reduces the achievable throughput ${\mathcal{R}}_{uk}^i$ in Eq. \eqref{eq:throughput} while the available resource is limited, i.e., bandwidth ${\mathcal{B}}_{uk}$.

\section{Conclusions}
\label{Conclusions}

In this paper, we have studied the mobile cellular networks in 6GBNs where one or more GBSs are inactive for energy-saving purposes. Moreover, we developed a novel Deep Q-learning-based algorithm that jointly optimizes the antenna tilt, power adaptation, MUs' RSRP, and MUs' throughput requirements to maximize the sum rate with a given number of served MUs satisfied. The proposed method not only solves the problem efficiently under dynamic and uncertain environments but also deals with the high dimensional state space and action spaces. The extensive simulation results showed that our proposed algorithm provides a big gain with different numbers of MUs and distance between GBS $\rightarrow$ MU values. 

\balance
\bibliographystyle{IEEEtran}
\bibliography{IEEEfull}
\end{document}